\documentclass[10pt,conference]{IEEEtran}
\IEEEoverridecommandlockouts
\usepackage{cite}
\usepackage{amsmath,amssymb,amsfonts}
\usepackage{graphicx}
\usepackage{textcomp}
\usepackage{xcolor}
\usepackage{subcaption}
\usepackage{algorithm}
\usepackage{algpseudocode}

\def\BibTeX{{\rm B\kern-.05em{\sc i\kern-.025em b}\kern-.08em
    T\kern-.1667em\lower.7ex\hbox{E}\kern-.125emX}}

\usepackage[acronym,nopostdot,  style=super, nonumberlist, toc]{glossaries}
\newacronym[plural=HITLs,firstplural=human-in-the-loop (HITL)]{HITL}{HITL}{human-in-the-loop}
\newacronym[plural=CPSs,firstplural=cyber-physical systems (CPS)]{CPS}{CPS}{cyber-physical system}
\newacronym[]{TSN}{TSN}{time sensitive networking}
\newacronym[]{FIFO}{FIFO}{first-in, first-out}
\newacronym[]{MCS}{MCS}{modulation coding scheme}
\newacronym[]{CDF}{CDF}{cumulative density function}
\newacronym[]{PDF}{PDF}{probability density function}
\newacronym[]{CCDF}{CCDF}{complementary cumulative density function}
\newacronym[]{AQM}{AQM}{active queue management}
\newacronym[]{DVP}{DVP}{delay violation probability}
\newacronym[]{SINR}{SINR}{signal to interference plus noise ratio}

\usepackage{amsmath,amsfonts} 
\usepackage{graphicx}

\newcommand*{\expe}{\mathbb{E}}
\DeclareMathOperator*{\argmax}{arg\,max}

\IEEEoverridecommandlockouts\IEEEpubid{\makebox[\columnwidth]{978-1-979-8-3503-1090-0/23/\$31.00~\copyright~2023 European Union \hfill} \hspace{\columnsep}\makebox[\columnwidth]{ }}

\begin{document}

\title{Active Queue Management with Data-Driven Delay Violation Probability Predictors\\
\thanks{This work was supported by the European Commission through the H2020 project DETERMINISTIC6G (Grant Agreement no. 101096504), in conjunction with Digital Futures research center.}
}

\author{\IEEEauthorblockN{Samie Mostafavi, Neelabhro Roy, György Dán and James Gross}
\IEEEauthorblockA{\text{KTH Royal Institute of Technology},
Stockholm, Sweden \\
\{ssmos, nroy, gyuri, jamesgr\}@kth.se}
}


\maketitle

\begin{abstract}
The increasing demand for latency-sensitive applications has necessitated the development of sophisticated algorithms that efficiently manage packets with end-to-end delay targets traversing the networked infrastructure.
Network components must consider minimizing the packets’ end-to-end delay violation probabilities (DVP) as a guiding principle throughout the transmission path to ensure timely deliveries.
Active queue management (AQM) schemes are commonly used to mitigate congestion by dropping packets and controlling queuing delay.
Today's established AQM schemes are threshold-driven, identifying congestion and trigger packet dropping using a predefined criteria which is unaware of packets' DVPs.
In this work, we propose a novel framework, Delta, that combines end-to-end delay characterization with AQM for minimizing DVP.
In a queuing theoretic environment, we show that such a policy is feasible by utilizing a data-driven approach to predict the queued packets' DVPs.
That enables Delta AQM to effectively handle links with
arbitrary stationary service time processes.
The implementation is described in detail, and its performance is evaluated and compared with state of the art AQM algorithms.
Our results show the Delta outperforms current AQM schemes substantially, in particular in scenarios where high reliability, i.e. high quantiles of the tail latency distribution, are of interest.
\end{abstract}


\section{Introduction}

\Glspl*{CPS} integrate computing and communication elements and physical processes, enabling intelligent control and monitoring of physical entities ~\cite{Lee_2008_cps}.
With the rise of \gls*{CPS}, \gls*{HITL} applications have emerged where humans interact with \gls*{CPS} to provide decision-making, supervision, and intervention capabilities. 
These applications range from augmented reality systems to autonomous vehicles and industrial automation.

Real-time applications, such as \gls*{CPS} and \gls*{HITL}, have stringent delay requirements, often expressed in terms of a target delay that should not be exceeded more than with a certain probability. 
This probability, known as the \gls*{DVP}, represents the likelihood that a packet will not reach its destination before the specified deadline~\cite{chang2000performance}.
For example, \gls*{HITL} applications typically have a delay target around 100 ms with a DVP of 0.99 to 0.999~\cite{sharma2023deterministic,schulz_latency_2017}.

Existing approaches to meeting such stringent latency requirements fall into two main groups. On the one hand, there is a large body of work on the characterization of the end-to-end latency distribution in computer networks, with queuing theory as a fundamental tool.
Deterministic network calculus approaches aim to analyze the worst-case network latency by considering specific service and arrival processes. 
Subsequently, stochastic network calculus emerged, focusing on deriving stochastic bounds on network latency.
However, these approaches encounter limitations as they rely on assumptions regarding the service process and apply only to service processes that are independent and identically distributed (i.i.d.) \cite{jiang2008stochastic,fidler2014guide,jaya_2020}.  
These limitations constrain their applicability in practical scenarios for \gls*{DVP} optimization.

The apparent limitations of analytical approaches have motivated recent interest  in data-driven end-to-end latency distribution prediction schemes~\cite{sawabeDelayJitterModeling2022,samie_sec_2021} and a renewed interest for  \gls*{AQM} ~\cite{admas_aqm_2012}. AQM drops packets from buffers before they would overflow, to prevent excessive end-to-end delays due to long queues.
Existing \gls*{AQM} schemes are threshold-driven, using predefined criteria to identify congestion and trigger packet dropping, and hence they cannot incorporate application-layer end-to-end delay requirements in their dropping decision.
Thus, the \gls*{AQM} dropping criteria do not change in response to changing latency requirements, instead tedious parameter
tuning is needed. Clearly, for making the large scale deployment of latency sensitive applications, there is a need for a methodological approach for achieving a target DVP in an automated manner. 

In this work, we propose a framework that combines end-to-end delay characterization with \gls*{AQM} for minimizing \gls*{DVP}.
The proposed framework makes it possible to  incorporate transient \gls*{DVP} predictions  in the AQM packet dropping decisions. Or main contributions are as follows.

\begin{itemize}
    \item We propose of a novel \gls*{AQM} framework, Delta, which effectively minimizes end-to-end delay violations by incorporating \gls*{DVP} predictions in packet dropping decisions.
    \item We make use of a data-driven approach for \gls*{DVP} prediction, enabling Delta \gls*{AQM} to effectively handle links with arbitrary stationary service time processes.
    \item We evaluate Delta \gls*{AQM} in a representative queueing environment and show that it outperforms state of the art AQM schemes.
\end{itemize}

\subsection{Related Work}
Active queue management works such as ~\cite{pan_pie_2013,codel_2012} aim to solve the bufferbloat problem, i.e., queues in the network causing high latency and delay jitter. 
Their proposed schemes PIE and CoDel aim to address this problem by limiting the average queueing latency to a reference value; and by using the minimum rather than average as the queue measure, simplified single-state variable tracking of minimum, and by using queue-sojourn time. 
There have been some recent works such as \cite{kim_deep_2021}, which utilize a deep reinforcement learning based architectures for queue management. 
The authors show performance gains over other deep-Q-networks (DQN) based works. 
This work utilizes a scaling factor in their reward function in order to arrive at the trade-off between queueing delay and the throughput. 
Since different applications would need different scaling factors, this approach does not generalize well and is heavily dependent on fine-tuning the reward functions.
Another line of research involves works which aim to minimize the end-to-end latency in queues. 
Liu \textit{et al.} in \cite{9686681} propose a method towards end-to-end congestion control and queue management, which aims to reduce the jitters and make the end-to-end latency deterministic. 
Building on top of this, authors in ~\cite{kar_paqman_2022} formulate a Semi-Markov Decision Process (SMDP) to obtain an optimal packet dropping policy. They realize this by assuming a probabilistic model for the flow rates and the de-queueing pattern. 
Their work focuses on characterizing the load, emanating out of their model of choice where packet data arrivals are given by gamma distributed inter-arrival times. 
Additionally, they assume the service times to be exponentially distributed which may not generalize well.

\section{System Model and Problem Statement}

In this study, we assume a single \gls*{FIFO} queue with a deterministic arrival process.
We consider such a simple model for exposing the problem and the proposed solution.
We discuss the extension of the proposed solution to multiple queues in Section \ref{sec:conc}.
\begin{figure}
  \begin{center}
    \includegraphics[width=0.9\linewidth]{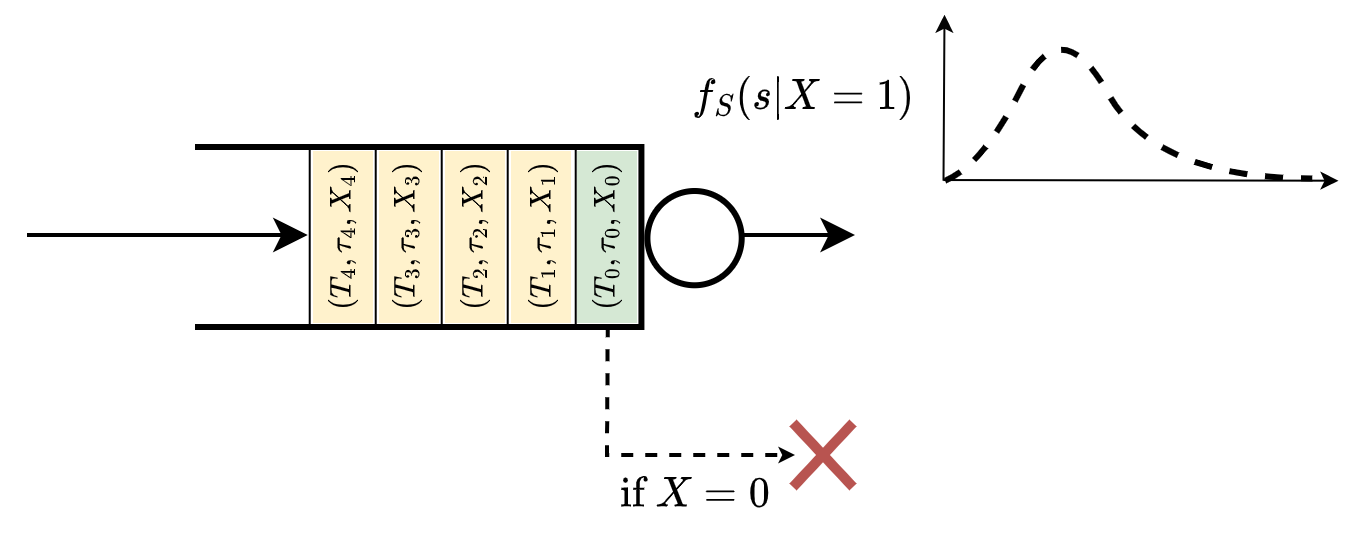}
  \end{center}
  \caption{Queueing model with \gls{AQM}}
  \label{fig:queue_model}
\end{figure}
Packets are generated at the start node and queued until the server becomes available and can process the corresponding packet.
Let $M$ = $\{1,...,m\}$ denote a set of consecutive time sensitive packets traversing the link.
We denote by $T_i$ the arrival time and by $\tau_{i}$ the target delay of packet $i \in M$.
We denote by $W_{i}$ the queueing time of packet $i$, by $S_{i}$ its service time, and by $Y_{i} = W_{i} + S_{i}$ the resulting sojourn time.
The server can decide to mark a queued packet to be dropped prior to service, denoted by $X_i=0$, in which case the packet has service time $S_i=\infty$ as shown in Figure \ref{fig:queue_model}.
A packet not marked for dropping, i.e., $X_i=1$, will enter the server and will experience service time $S_i \sim f_S(s \mid X=1)$ with a general stationary process.
The selection of a general distribution for the service delay process stems from our objective to replicate the stochastic delay observed in network links, particularly prevalent in wireless networks.
Therefore, the probability density function of the service time is denoted by $f_S(s \mid X)$ where the variable $X \in \{0,1\}$ models the capability of the server refusing to process the packet, i.e. to drop the packet.
By default, all packets are assigned $X_i = 1$ on their arrival.

\textbf{Problem Statement}: 
Our optimization objective is the fraction of successfully processed packets after the $m$-th packet has been processed. 
It is denoted by
\begin{equation} \label{eq:performance}
R_M = \frac{\sum_{i=1}^{m} \mathfrak{I}\left[Y_{i} \leq \tau_{i} \right]}{m},
\end{equation}
where $\mathfrak{I}\left[\cdot \right]$ is the indicator function equalling $1$ if the argument is true and $0$ otherwise.
Formally, an \gls{AQM} scheme provides a mapping from the currently backlogged packets to a dropping decision vector, denoted by $\pi: \mathcal{D}_n \rightarrow \mathcal{X}_{n}$.
The goal is to find the \gls{AQM} policy $\pi^*$ that maximizes the fraction of successful packets $R_M$ by deciding about dropping some of the packets of the stream, denoted by the dropping variable $\mathbf{X} := \{ X_{i} \}^{m}_{i=1}$.

\section{Approach}
To solve this problem, we strive to find the dropping vector that maximizes the expected outcome limited to the $n$ packets in the queue $( N \subset M )$ denoted by
\begin{equation} \label{eq:opt1}
x^{\ast} = \argmax_{x \in \mathcal{X}_n} \left( \expe\left[ R_N|\mathbf{X}=x\right] \right),
\end{equation}
where $\mathbf{X}$ represents the dropping vector.
Packet drops can be applied frequently to increase the effect, particularly if the state of the queue changes rapidly.
In order to describe the queue state at dropping time $t$, we use the notion of the remaining delay budget of each packet denoted by
\begin{equation}
    \delta_{i,t} = \max(\tau_{i} - (t-T_i),0).
\end{equation}
Hence, we denote the queue state at time $t$ by vector $\Delta_{t} := \{ \delta_{i,t} \}^{n}_{i=1}$.
Remaining delay budget is considered a decisive attribute of the packets during the decision rounds.
For instance, if the remaining delay budget of a packet is very small, the \gls{AQM} algorithm can decide not to serve it, so the successors of the packet face less waiting time.

Our proposed AQM scheme \textit{Delta} essentially chooses at every decision round the dropping vector that maximizes the expectation of successfully transmitted packets from the queue state vector.
We solve Equation \ref{eq:opt1} by finding the optimization function as
\begin{equation} \label{eq:opt}
\expe\left[R_{N}|\mathbf{X}=x\right] =  \frac{\sum_{i=1}^{n} \expe\left[\mathfrak{I} \left[ Y_{i} \leq \tau_{i} \right] |\mathbf{X}=x\right] }{n}.
\end{equation}
The right-hand side expectation term in Equation \ref{eq:opt} essentially describes the probability of packet $i$ finishing before its deadline $\tau_{i}$, given the dropping vector is applied.
We call this expectation the packet's success probability $\psi_{i,x}$ and it can be obtained by calculating the complementary of the packet's \gls{DVP} denoted by $\varphi_{i,x}$.
\begin{equation}
\psi_{i,x} = 1 - \varphi_{i,x} = 1 - \expe\left[\mathfrak{I} \left[ Y_{i} > \tau_{i} \right] | \mathbf{X}=x \right].
\end{equation}
The problem is narrowed down to finding the \gls{DVP} for each packet in the queue.
Assuming such estimation is feasible, a decision tree is formed to represent the set of all possible dropping vectors and their outcomes.
\begin{figure}
    \begin{center}
    \includegraphics[width=0.7\linewidth]{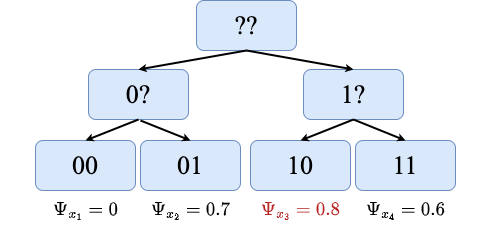}
    \end{center}
    \caption{The decision tree of dropping packets in a queue of length 2 where 0 means drop and 1 means pass.}
    \label{fig:decisiontree}
\end{figure}
As shown in Figure \ref{fig:decisiontree}, the decision tree is solved by traversing the tree from the root to the leaf nodes, evaluating the objective function mentioned in Equation \ref{eq:opt} associated with each path through the tree. 
The optimal solution is the path through the tree that results in the highest $\Psi_{x_j} = \sum_{i=1}^{n} \psi_{i,x_j}$.
\begin{figure}
  \begin{center}
    \includegraphics[width=8cm]{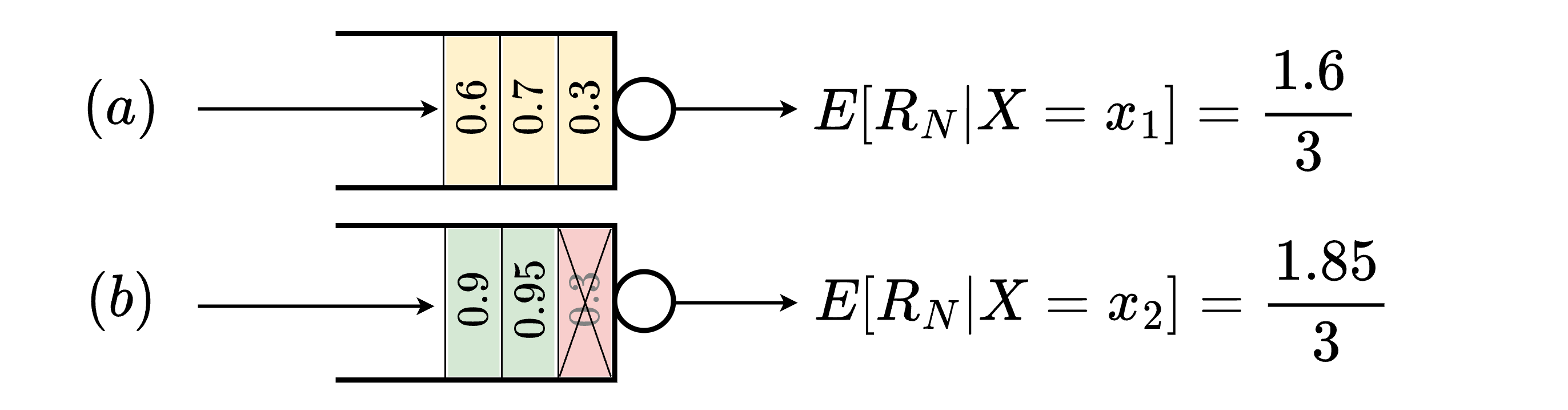}
  \end{center}
  \caption{An example on calculating the objective function for the dropping vectors $x_1$ and $x_2$ which consider if packets in the queue should be dropped.}
  \label{fig:interventionexample}
\end{figure}
In a more detailed example shown in Figure \ref{fig:interventionexample}, the dropping vectors $x_1$ and $x_2$ are compared to decide if the first packet should be dropped or not.
The calculated objective function in b is higher, so it is favored.

To estimate the packets' \glspl{DVP}, we resort to the approach described in our previous work ~\cite{samie_sec_2021}.
At its core, we estimate the \gls{PDF} of the remaining latency $f_{Z}$ for any packet, from the number of its predecessors in the queue $X$.
For the $i$th packet in the queue at time $t$, \Gls{DVP} is obtained by calculating the remaining latency's \gls{CCDF} at the remaining delay budget $\delta_{t,i}$ denoted by
\begin{equation}
  \varphi_{i,x} = f_{Z}(z > \delta_{t,i} \mid X).
\end{equation}
Considering the dropping vector $x$ chosen by Delta \gls{AQM}, we must incorporate the number of effective predecessors, or the predecessors that are not going to be dropped by $X = \textstyle\sum_{j=1}^{i}x_j$.

We utilize a machine learning approach, namely mixture density networks, for latency density estimation for any queued packet at any time in a queuing system.
In mixture density networks, the parameters of a parametric density function e.g. a Gaussian mixture model, are controlled by a fully connected neural network.
Therefore, we approximate $\varphi_{i,x}$ by the parametric density function $\hat{p}_{\theta}$ as
\begin{equation}
      \varphi_{i,x} \approx \hat{p}_{\theta}(z > \delta_{i,t} \mid X = \textstyle\sum_{j=1}^{i}x_j ).
\end{equation}
To train the neural network, a dataset must be formed by recording the number of predecessors when a packet enters the queue and the end-to-end latency.

The description of our AQM policy Delta with the conditional density function is described in Algorithm \ref{alg:algorithm}.
\begin{algorithm}
\caption{Delta}
\label{alg:algorithm}
    \begin{algorithmic}[1]
    \Statex
    \Function{$\mu$}{$\Delta_t$}
        \State {$\Psi_{max}$,$\hat{x}$ $\gets$ {$0$}}
        \For{$x \in \mathcal{X}_n$}
            \State {$\Psi_x$ $\gets$ {$0$}}
            \For{$i \in N$}
                \State $\psi_{i,x} \gets {\hat{p}_{\theta}(z \leq \delta_{i,t} \mid X = \textstyle\sum_{j=1}^{i}x_j )}$
                \State $\Psi_x \gets \Psi_x + \psi_{i,x}$
            \EndFor
            \If{$\Psi_x > \Psi_{max}$}
                \State $\Psi_{max} \gets \Psi_x$
                \State $\hat{x} \gets x$
            \EndIf
        \EndFor
        \State \Return {$\hat{x}$}
    \EndFunction
\end{algorithmic}
\end{algorithm}
The time complexity of Delta is O($2^{N}$), which is relatively high and could pose a challenge in practical applications. 
This complexity indicates that the algorithm's computational requirements increase exponentially with the number of inputs, making it necessary to use a limit on the queue length. 
Parallel processing techniques can be utilized to reduce the algorithm's run time. 
These findings suggest room for further investigation into improving the algorithm's efficiency in future research.
\section{Evaluation}

This section comprehensively analyzes the proposed Delta \gls{AQM} scheme in a simulated queuing system as shown in Figure \ref{fig:queue_model} \footnote{The reproducible results could be found at: https://github.com/samiemostafavi/delta-queue-management}.
Firstly, we present a comparative assessment between Delta \gls{AQM} and two established and advanced schemes: CoDel, widely implemented in network infrastructures \cite{codel_2012}, and DeepQ \cite{kim_deep_2021}, a state-of-the-art method based on Deep Reinforcement Learning.
Furthermore, we investigate Delta \gls{AQM}'s sensitivities to different factors in-depth. 
We assess its efficacy in diverse scenarios, such as situations with various delay targets and utilization factors.
The utilization factor is determined by dividing the packet arrival rate by the average service rate.
Additionally, we analyze the performance of Delta \gls{AQM} in scenarios where there is a mismatch between the \gls{DVP} predictor's learned behavior about the link and the actual network conditions, providing insights into its adaptability.

In simulations, we model the queue's service delay process by a Gamma distribution where, by default, the concentration is set to $5$ and the rate to $0.5$ resulting in $0.1$ as the average service rate.
The tasks arrival process is assumed to be deterministic to represent sensory data traversing the network link.
Tasks arrive at the queue carrying a target delay which is assumed to be constant for all tasks in each simulation.
We analyze the performance over a range of target delay values.
This range is defined to cover the delay target values that are close or far compared to the average end-to-end delay in the no-aqm simulation. 
For instance, in all figures, the $0.8$ target delay corresponds to the $0.8$ quantile of the end-to-end delay of the no-aqm simulation.
In all benchmarks we compared the \gls{AQM} schemes by the failed tasks ratio introduced in Equation \ref{eq:performance}.
We set the number of completed tasks $m$ in the simulations to at least $10^7$ which is large enough to mitigate the performance variance even for large target delays.
In order to prevent excessively long processing times during simulations, we constrained the AQM to focus on the initial 15 packets within the queue.

We include the no-AQM system performance and offline-optimum policy performance in the comparisons.
The offline-optimum policy already knows the end-to-end delay of the tasks in the queue.
At every decision round, it drops the head packet if its end-to-end delay exceeds the delay target.
Thus, every packet being served will make the deadline.
Due to access to the delay information, the offline-optimum \gls{AQM} scheme's performance is expected to be superior to all others.

\begin{figure}
\centering
\begin{subfigure}{1\linewidth}
    \begin{center}
        \includegraphics[width=1\linewidth]{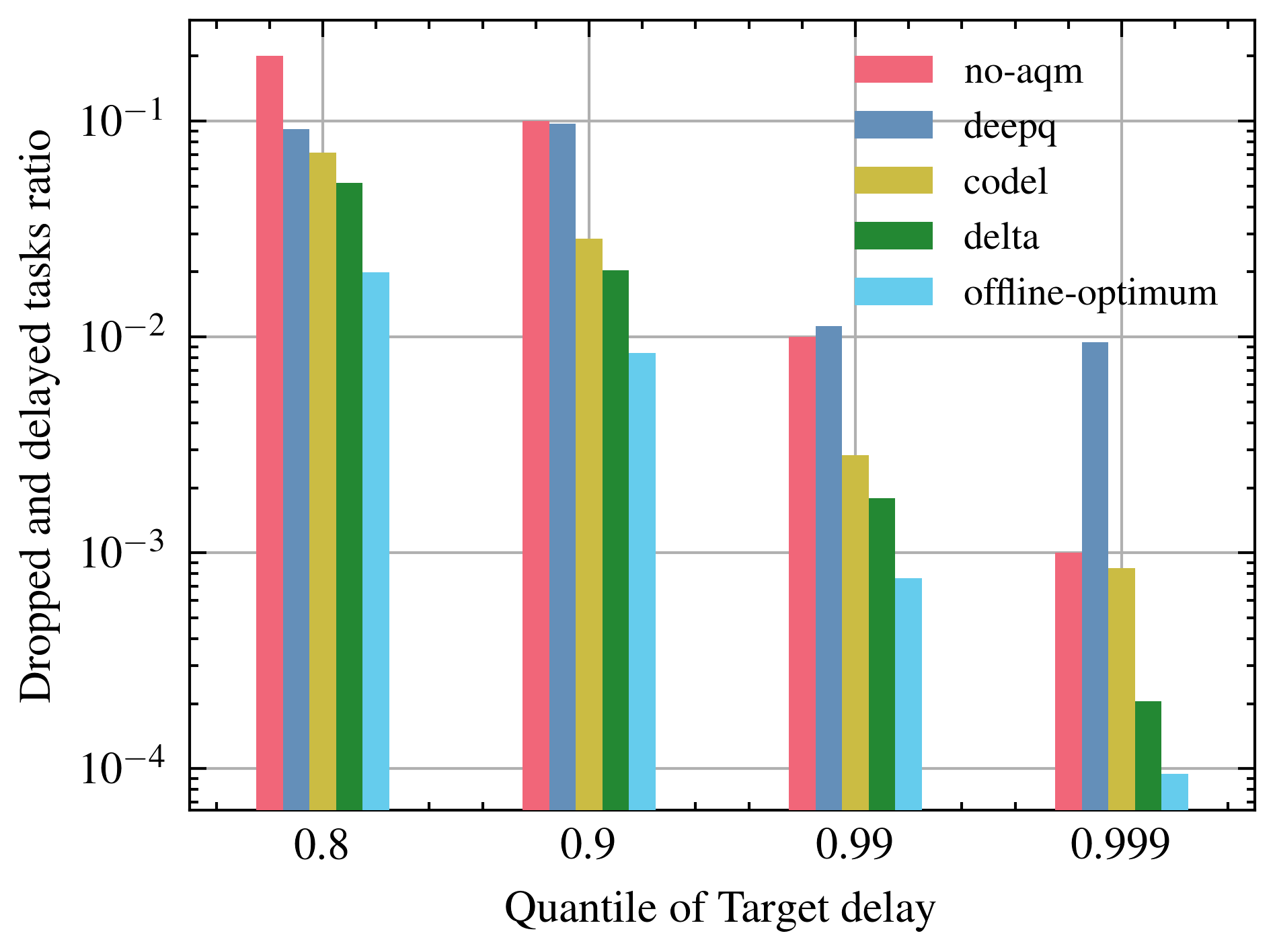}
    \end{center}
    \caption{Utilization factor: $91.6\%$}
    \label{result1}
\end{subfigure}
\begin{subfigure}{1\linewidth}
  \begin{center}
        \includegraphics[width=1\linewidth]{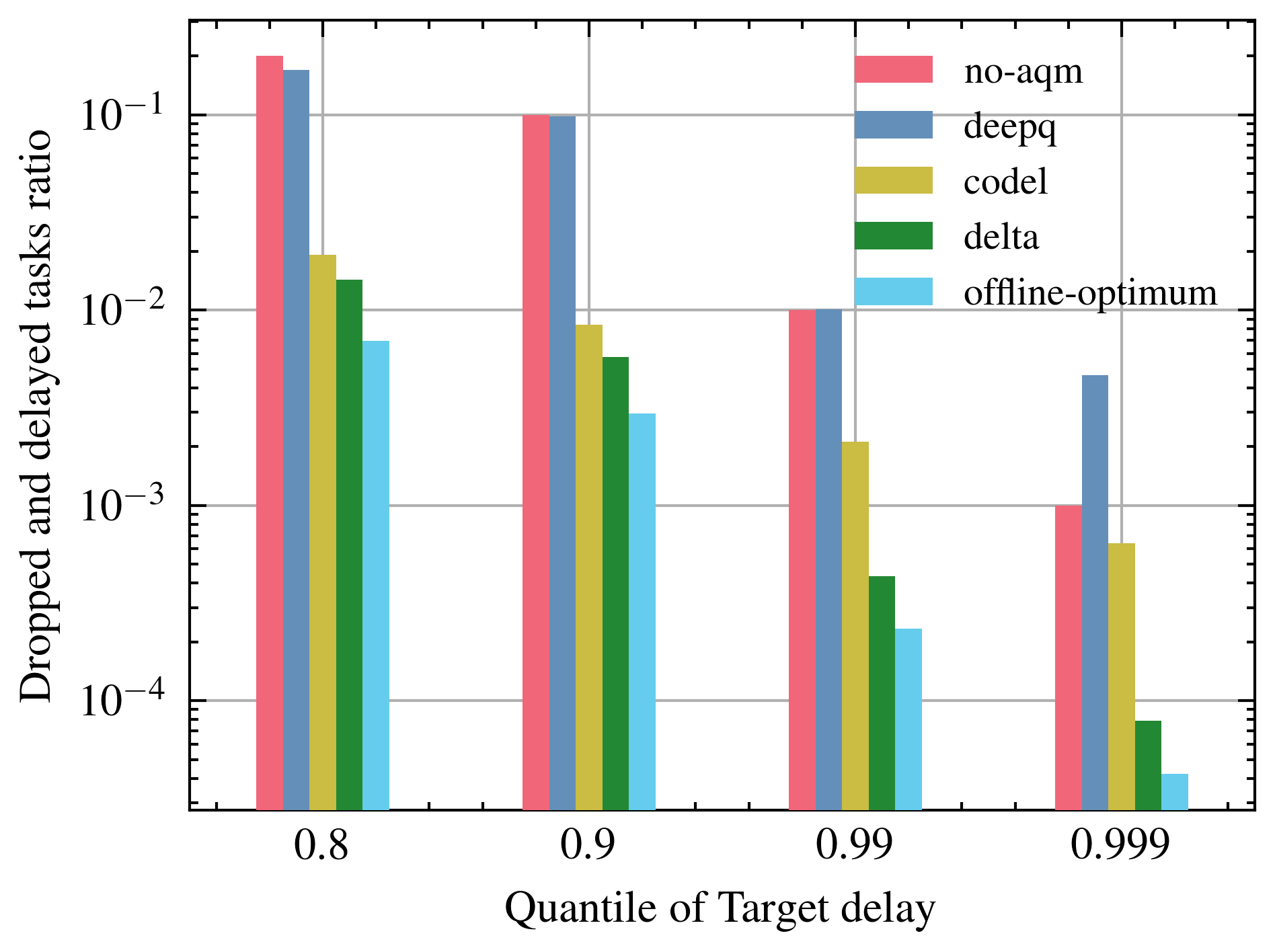}
    \end{center}
    \caption{Utilization factor: $96.7\%$}
    \label{result2}
\end{subfigure}
\caption{Comparing different \gls{AQM} schemes}
\end{figure}

The evaluation starts by comparing our proposed approach with CoDel \cite{codel_2012}, DeepQ \cite{kim_deep_2021}, the offline-optimum and the case with no-AQM.
CoDel, as mentioned earlier, operates based on predefined criteria and thresholds to detect congestion and initiate packet dropping, which are determined by interval time and target delay. 
In our evaluation, we compare the performance of the \gls{AQM} schemes across four different target delay values, varying from those close to the average no-AQM delay quantiles to those significantly further. 
Notably, for each target delay, it was essential to readjust the CoDel parameters; otherwise, the scheme's performance exhibited a notable degradation. 
The same requirement applied to DeepQ, necessitating a retraining phase to adapt its policy parameters for each target delay. 
In contrast, our proposed Delta scheme demonstrates an advantageous characteristic, as it does not require any parameter tuning or retraining when the packet's delay targets change. 
The same Delta model remained applicable and effective across all four target delays without modifications.
We demonstrate this in Figure \ref{result1} and \ref{result2} where we increase the utilization factor moving from Figure \ref{result1} to Figure \ref{result2}.
As can be observed from the figures, our proposed approach consistently performs better than the other approaches, barring the offline-optimum.
On moving from Figure \ref{result1} to Figure \ref{result2} we can observe an improvement in performance for the algorithms (except no-AQM) in terms of fewer task failures. 

\begin{figure}
    \begin{center}
        \includegraphics[width=\linewidth]{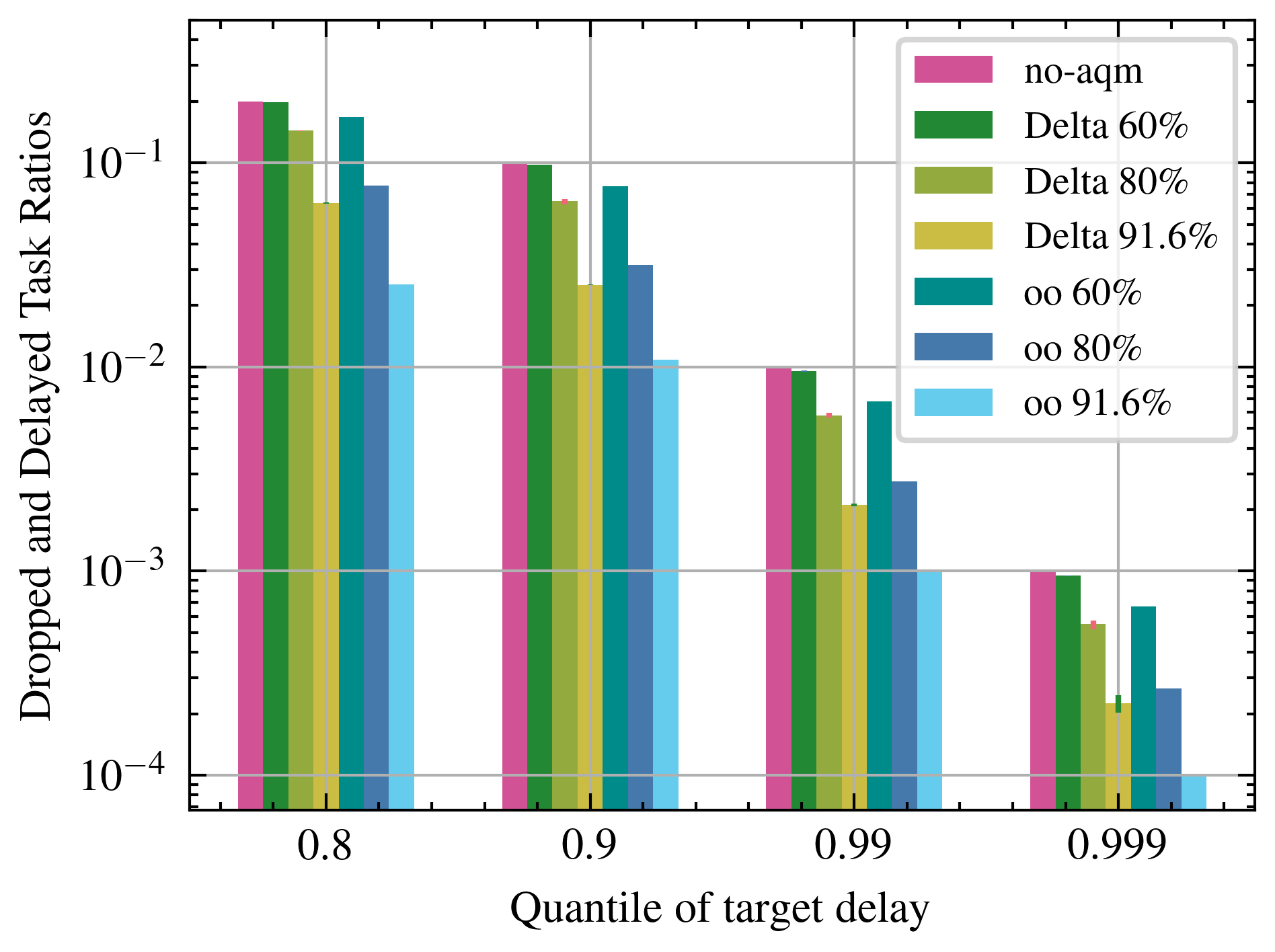}
        \vspace{0.1cm}
    \end{center}
    \caption{Effect of sparsely populated queues}
    \label{80Util}
\end{figure}

We extend the analysis of utilization factors by comparing 91.6\% and 96.7\% from Figure \ref{result1} with 80.7\% and 60\% additionally in Figure \ref{80Util}.
"Delta 60" in Figure \ref{80Util} corresponds to a utilization factor of 60\% and similarly "Delta 80" corresponds to a utilization factor of 80.7\% and finally "Delta 91.6" in Figure \ref{80Util} corresponds to a utilization factor of 91.6\%.
The nomenclature for the offline optimum schemes follows a similar trend, being abbreviated as "oo".
Here we can observe that as the utilization factor drops appreciably to 60\%, Delta's performance is almost similar to that of no-AQM.
As the utilization factor increases to 80.7\%, the performance improves and finally at 91.6\%, the performance is significantly better than that of no-AQM. 
The offline optimum approach too follows a similar trend as Delta, approaching no-AQM closely for a utilization of 60\% and then gradually improving as the utilization factor increases.
All these findings can be attributed to how, for sparsely populated queues, there are more drops in terms of the total number of packets in the queue at any time.
It is also worth noting that the target delay corresponding to the 0.8th quantile with the utilization factor of $80.7\%$ is different from the one with utilization factor of $91.6\%$.

\begin{figure}
    \begin{center}
        \includegraphics[width=\linewidth]{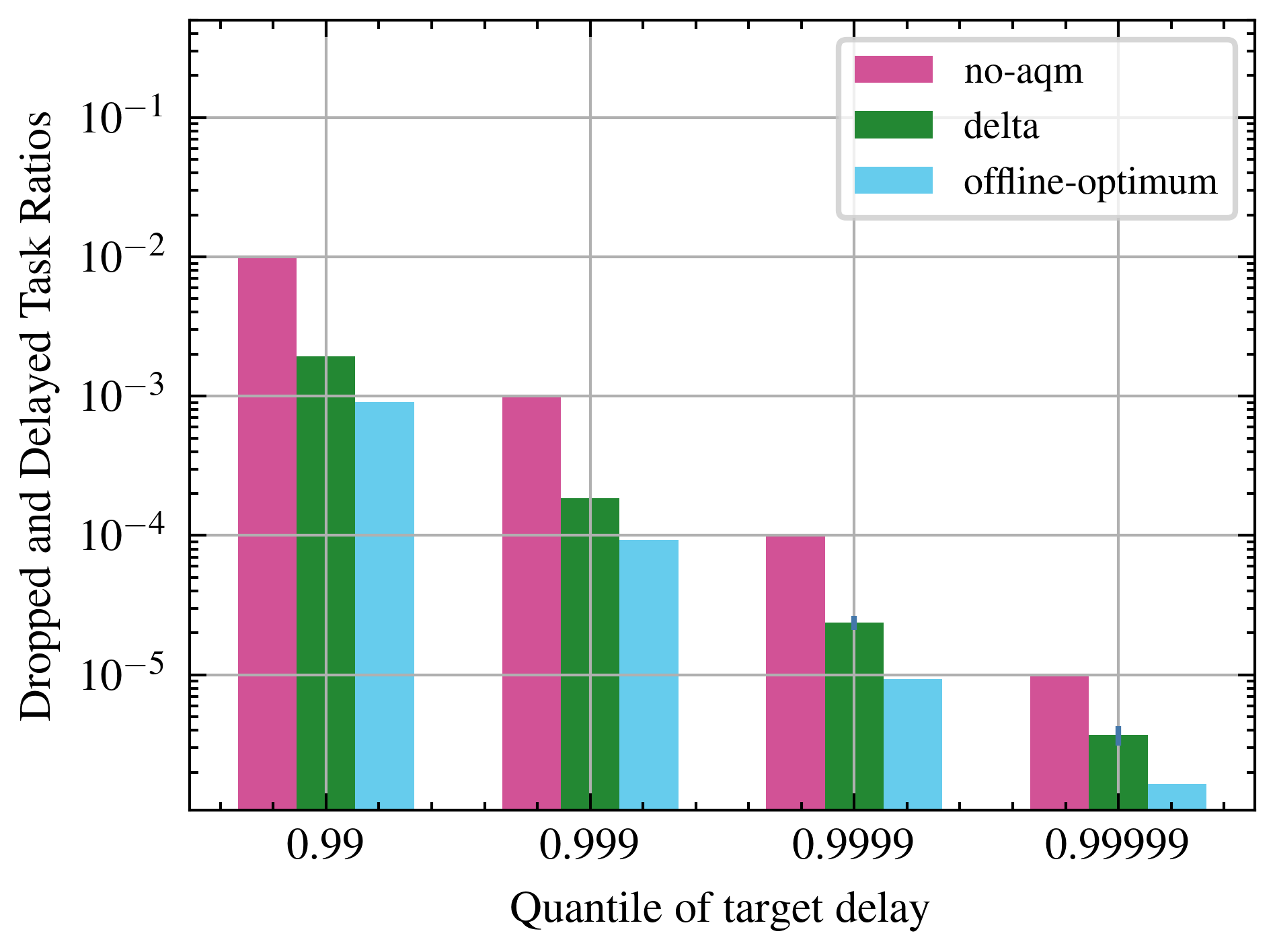}
        \vspace{0.1cm}
    \end{center}
    \caption{Analysis of higher quantiles of target delay, utilization factor: $91.6\%$}
    \label{59sbench}
\end{figure}

Figure \ref{59sbench} extends our analysis for the implementation with the utililzation factor of 91.6\% to higher quantiles of target delays.
In this plot, we can additionally observe and compare no-AQM, our proposed approach Delta and the offline optimum for the 0.9999th and 0.99999th quantile of target delays.
It is interesting to note here that across the various quantiles of target delays being analyzed, the offline optimum performs at least 10 times better than the no-AQM scheme in terms of task failure ratios. Moreover, our proposed solution approaches the offline-optimum closely.

\begin{figure}
    \begin{center}
        \includegraphics[width=\linewidth]{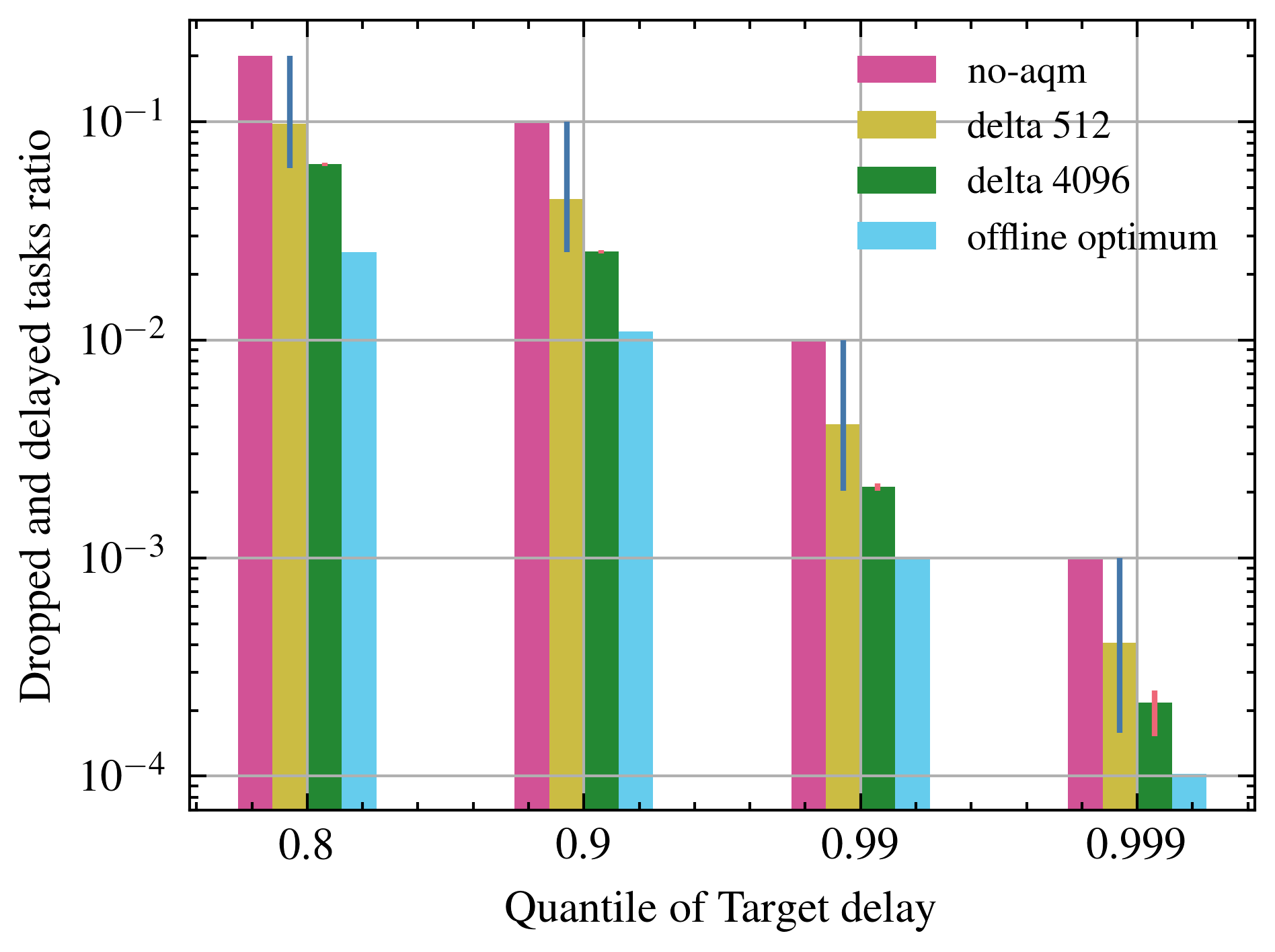}
        \vspace{0.1cm}
    \end{center}
    \caption{The effect of using a poorly trained DVP predictor on the performance of the Delta AQM, utilization factor: $90.6\%$}
    \label{mismatch}
\end{figure}

Next, we analyze the performance when the DVP predictor is trained over a varying number of samples in Figure \ref{mismatch}.
The label "delta 512" denotes the case where our proposed algorithm is just trained over 512 samples.
This is however repeated over multiple iterations and the average of the runs is shown, along with the minimum and maximum range of the values obtained (by the blue marker).
Here we can observe that while the average of the different runs has a better performance (in terms of task failures) than the case with no AQM, in some runs, the performance equals that of "No AQM".
The training with just 512 samples can thus be concluded as inadequate.
On the other hand, when trained over 4096 samples, the performance is consistently better than the case with no AQM and also the case with "delta 512", showing that increasing the number of training samples has a positive impact.

Finally, we evaluate our approach over cases where our predictor is exposed to differing Gamma service processes against a different service process it was trained over.
\begin{figure}
    \begin{center}
        \includegraphics[width=\linewidth]{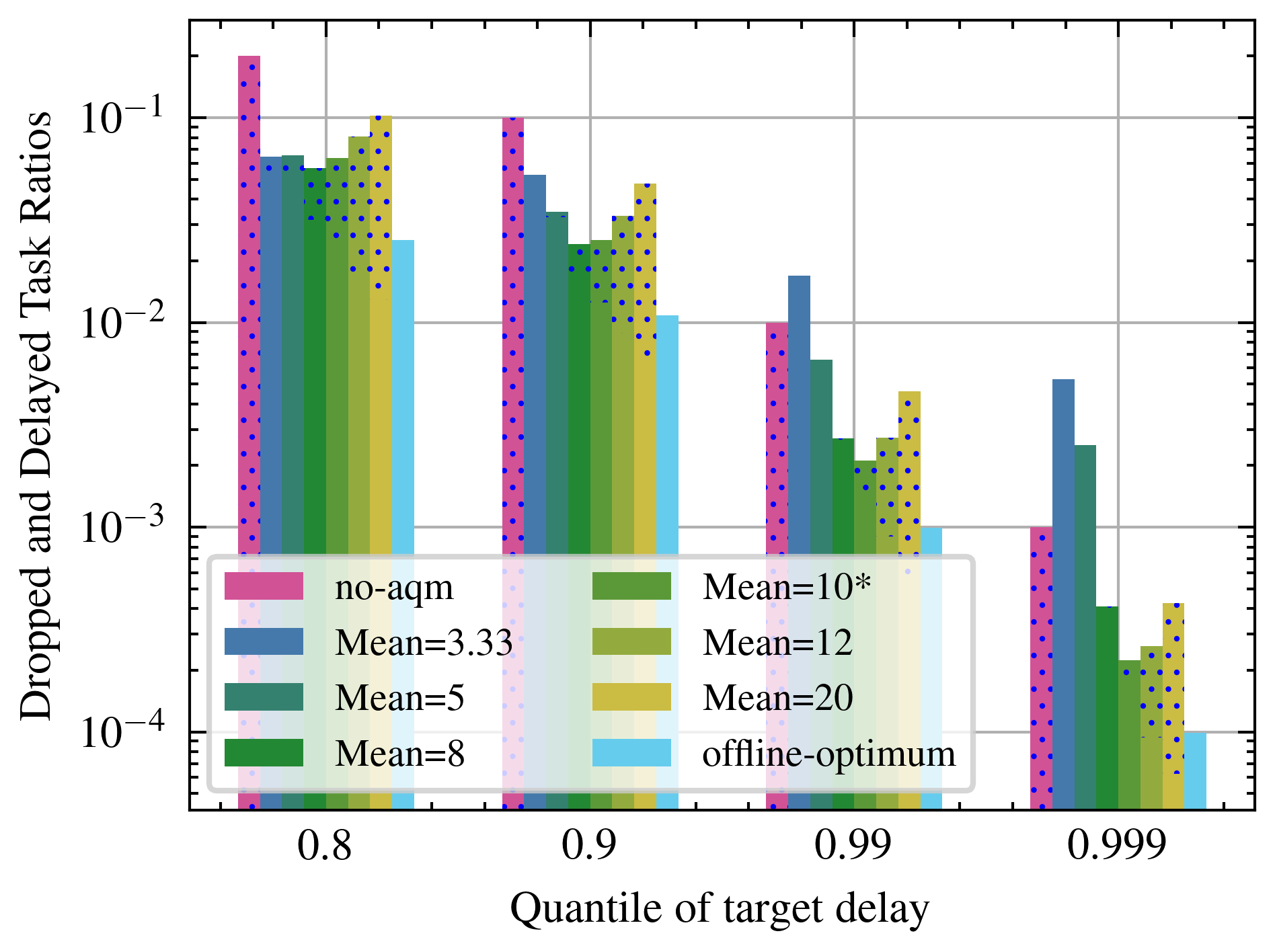}
        \vspace{0.1cm}
    \end{center}
    \caption{The effect of testing against varying Gamma rates on the performance of the Delta AQM, utilization factor: $90.6\%$}
    \label{mismatch_main}
\end{figure}
Figure \ref{mismatch_main} shows how different algorithms (and our proposed algorithm, tested in varying conditions) perform against each other with respect to the failed task ratios, observed across specific target delays.
While our proposed algorithm has undergone the same training process here, it has been tested across different service processes which correspondingly have different means and variances. 
The dots in the bars represent the delayed packet ratio out of the total tasks. 
The transparent region (non-dotted) represents the dropped packet ratio.

We can observe that the least number of failed tasks correspond to the case where the predictor is tested against a service process which follows the same delay distribution as the training process. 
The Gamma rate of the service process is then varied, while also proportionately changing the packet arrival rate, so as to maintain the same utilization factor in the benchmark. 
We compare our proposed approach against a case where no-AQM is performed and a case where the queue manager has all the information about incoming packets apriori which we label as the offline-optimum. 
For the case where our proposed approach is trained and tested over the same service process (denoted by Mean = 10*), we can observe that the performance is consistently better than no-AQM.
This also holds true when we change the mean of the service process by +-2 while testing (denoted by Mean = 8 and 12). 
Interestingly, when we test Delta over a significantly different service process (Mean= 3.33) than what we trained over, it performs worse than no-AQM case.
As we move higher from our 10* delta scheme, towards a higher mean of the service process, we observe a significantly higher proportion of delayed tasks owing to the higher service mean, as most tasks can be accommodated in principle, but at the cost of them exceeding their delay bounds.
As we move lower from our 10* delta AQM scheme, towards a lower service mean, we observe that delayed packets are significantly lesser than the dropped packets. 
This is due to the fact that when the service process has a lower mean, there are more failed tasks, most of which cannot be accommodated without them being dropped. 
The general observation however remains that on testing across varying Gamma rates, there are more task failures.
\section{Conclusions} \label{sec:conc}
This paper presented Delta, a novel AQM scheme that effectively minimizes end-to-end delay violations by incorporating DVP predictions in dropping decisions. 
By utilizing a data-driven DVP predictor, Delta AQM offers a promising solution for links with general stationery service time processes.
Numerical evaluations in a queuing theoretic environment demonstrate that Delta performs consistently well across a wide range of delay targets without the need for modifications.
Comparative assessments highlight the advantages of Delta's parameter-free approach compared to other schemes that require parameter adjustments or retraining for each target delay.
Additionally, we conducted a thorough analysis of Delta AQM's sensitivity to the number of training samples, as well as its performance under mismatched scenarios where the characteristics of the link service process differ from the trained model. These investigations shed light on the sensitivities of Delta AQM, driven by the underlying machine learning algorithm.

In future work, an important direction of investigation will focus on assessing the efficacy and advantages of Delta AQM in multi-hop queues. 
In the multi-hop scenario, the utilization of the data-driven DVP predictor holds the potential to construct a more comprehensive probabilistic model of latency, enabling enhanced decision-making capabilities. 
This investigation will provide valuable insights into the applicability and benefits of Delta AQM in complex multi-hop network environments.



\bibliographystyle{ieeetr}
\bibliography{refs.bib}

\end{document}